\documentclass[aps,prl,reprint,showpacs,showkeys,superscriptaddress,floatfix]{revtex4-1}
\usepackage{multirow}
\usepackage{graphicx}
\usepackage{bm}
\usepackage{dcolumn}
\usepackage[colorlinks=true,linkcolor=blue,urlcolor=blue,citecolor=blue]{hyperref}
\usepackage{natbib}
\usepackage{amsmath}
\usepackage{times}
\usepackage{xcolor}
\usepackage{soul}
\usepackage[normalem]{ulem}

\preprint{1}
	
\begin{document}

\title{Anomalous lattice contraction and emergent electronic phases in Bi-doped Eu$_2$Ir$_2$O$_7$ }
\author{Prachi Telang}
\author{Kshiti Mishra}
\affiliation{Department of Physics, Indian Institute of Science Education and Research, Dr. Homi Bhabha Road, Pune 411 008, India}
\author{Giacomo Prando}
\affiliation{Dipartimento di Fisica, Università degli Studi di Pavia, I-27100 Pavia, Italia}

\author{A. K. Sood}
\affiliation{Department of Physics, Indian Institute of Sciences, Bangalore 560012, India}
 
 \author{Surjeet Singh}
\affiliation{Department of Physics, Indian Institute of Science Education and Research, Dr. Homi Bhabha Road, Pune 411 008, India}
\affiliation{Center for Energy Sciences, Indian Institute of Science Education and Research, Dr. Homi Bhabha Road, Pune 411 008, India}

\date{\today}

\begin{abstract}

We study the pyrochlore series (Eu$_{1-x}$Bi$_x$)$_2$Ir$_2$O$_7$ for $ 0 \leq x \leq 1$. We show that for small $x$, the lattice undergoes an anomalous contraction but the all-in/all-out and metal-to-insulator transitions remain robust, and the resistivity approaches a $1/T$ dependence at low-T, suggesting proximity to the Weyl semimetallic phase, as previously predicted theoretically. At the boundary between Eu$_2$Ir$_2$O$_7$ and Bi$_2$Ir$_2$O$_7$ a qualitatively different ground state emerges, which is characterized by its unusual metallic behavior and absence of magnetic ordering at least down to $0.02$ K.  

\end{abstract}


\maketitle
The $5d$ transition metal based oxides (TMOs) show intriguing behaviors lying outside the realm of strongly correlated systems \cite{Balents,Krempa2014,Cao2018}. A defining feature of these materials is their relatively large spin-orbit coupling $\lambda$ which competes with the on-site Coulomb repulsion $U$ to give rise to a range of novel correlated electronic phases, including quantum spin liquids, spin-orbit Mott insulators, Weyl semimetal and axion insulators \cite{Krempa2014, NakatsujiPr227, Wan}. In the iridates comprising Ir$^{4+}$ ($5d^5$) ions,  this interplay is further enriched as the crystal field interaction splits the $t_{2g}$ orbitals to yield an effective $J_{eff}  = 1/2$ moment, which forms the basis for studying a range of exotic quantum phenomena. 

Among iridates, the pyrochlores $A_2$Ir$_2$O$_7$  ($A:$ tri-positive rare-earth, Y or Bi ion) are of particular interest. The ground state of $A_2$Ir$_2$O$_7$ is sensitive to the choice of \textit{A}-cation, varying from insulating for $A =$ Gd$-$Lu and Y to metallic for $A =$ Pr and Bi \cite{Yanagishima, Matsuhira2011}. The intermediate members $A =$ Nd, Sm and Eu, on the other hand, show a distinct metal-to-insulator (MI) transition concurrent with the onset of \textit{all-in/all-out} (AIAO) magnetic ordering \cite{Disseler02, Sagayama, Donnerer,Guo}. For these intermediate members, Wan et al.  \cite{Wan} predicted a Weyl semimetallic (WSM) phase, which emerges in a narrow region of the $U - \lambda$ phase space when the Kramer's degeneracy of the quadratic band touching point  at $\Gamma$ is lifted due to breaking of the time reversal symmetry (TRS) in the AIAO state. This results in a pair of non-degenerate, linearly dispersing modes associated with the Weyl nodes of opposite chirality located on the Fermi energy ($\epsilon_F$). Following this pioneering work, there have been several theoretical predictions of novel topological phases in the pyrochlore iridates that can be realized by breaking either the time reversal or the cubic symmetry \cite{Krempa2013, Moon, Krempa2014, Savary, Yang1, Ueda}. However, the experimental realization of these phases has been limited so far.

Here, we study (Eu$_{1-x}$Bi$_x$)$_2$Ir$_2$O$_7$ for $0 \le x \le 1$. Both the end members of this series feature a non-magnetic $A^{3+}$-cation but are located in different regions of the $U-\lambda$ phase space. This allows one to span a wide region of this parameter space without disturbing the Ir$^{4+}$ valence state, which is key to realizing various non-trivial topological phases in these materials. While Eu$_2$Ir$_2$O$_7$ (EIO) is a candidate Weyl semimetal, Bi$_2$Ir$_2$O$_7$ (BIO) shows a metallic behavior with at least two magnetic phase transitions below T = 2 K, and an unconventional $T^{3/2}$ dependence of resistivity over a broad temperature range \cite{Qi,Baker}. We show that upon substituting large-size  Bi$^{3+}$  for Eu$^{3+}$, the lattice undergoes an anomalous contraction. Surprisingly, within this anomalous range the AIAO/MI transitions remain robust and become even sharper than for $x = 0$, and the resistivity approaches the 1/T dependence that was theoretically predicted for the WSM phase \cite{Hosur}. On the other hand, in the narrow range $ 0.035 < x < 0.1$, AIAO/MI transitions are strongly suppressed. Eventually, at the boundary $x = 0.1$, a novel ground state emerges which is characterized by an unusual metallic behavior and the absence of any long-range magnetic ordering down to at least 0.02 K. For $x > 0.1$, the resistivity behavior shows a surprisingly rich variation, suggesting the presence of interesting novel ground states. 

\begin{figure}[t!]
	\includegraphics[width =8.5 cm]{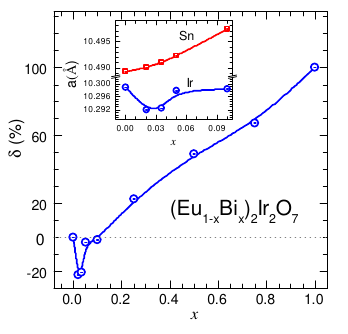}
	\caption{\label{lattice} $x$ dependence of the parameter $\delta$. The inset shows the variation of $a(x)$ in the range $0 \leq x \leq 0.1$ for (Eu$_{1-x}$Bi$_x$)$_2$Ir$_2$O$_7$ (Ir) and (Eu$_{1-x}$Bi$_x$)$_2$Sn$_2$O$_7$ (Sn).}
\end{figure}

The (Eu$_{1-x}$Bi$_x$)$_2$Ir$_2$O$_7$ ($x$ $=$ $0$, $0.02$, $0.035$, $0.05$, $0.1$, $0.25$, $0.50$, $0.75$, and $1$) samples were prepared using the solid-state method. The details of synthesis protocol are similar to that given in Ref. \cite{telang}. The lattice parameters were obtained using the Rietveld refinement done on very high-resolution synchrotron data (MCX at ELETTRA).  For comparison, we also prepared and studied homologous (Eu$_{1-x}$Bi$_x$)$_2$Sn$_2$O$_7$ ($0 \le x \le 0.1$) samples. Magnetic susceptibility ($\chi$) and resistivity ($\rho$) measurements were performed using a Physical Property Measurements System (Quantum Design). Muon spin spectroscopy ($\mu^{+}$SR) measurements were done at General Purpose Surface-muon (GPS) and Low Temperature Facility (LTF) spectrometers at the Paul Scherrer Institute (PSI), Switzerland.

The powder x-ray diffraction of the entire (Eu$_{1-x}$Bi$_x$)$_2$Ir$_2$O$_7$ series could be indexed based on the pyrochlore structure. In particular, we observe no change of symmetry or phase separation for any of the intermediate members in our high-resolution synchrotron data. In Fig. \ref{lattice}, we show the variation of lattice parameter $a$ and the deviation $\delta$, which is defined as $\delta(x) = (a - a_{0})/(a_{1} - a_{0}$); where $a_{0}$ and $a_{1}$ are the lattice parameters of EIO and BIO, respectively. In the range $x \leq 0.035$, $a$ exhibits an anomalous contraction with $\delta$ as large as $-21 \%$. On the contrary, in the homologous stannate series, doping with Bi leads to a regular lattice expansion as expected from the Vegard's law (see Fig. \ref{lattice} inset). Our results suggest a violation of the Vegard's law in the iridates series. Hereafter, we will refer to the negative lattice expansion region $0 \leq x \leq 0.035$ as \textit{anomalous}, the region $0.1 < x\leq 1$ where lattice expands normally as \textit{normal}, and the intermediate region $0.035 <x < 0.1$ as the \textit{crossover} region. 

\begin{figure}[t!]
	\includegraphics[width =8.3 cm]{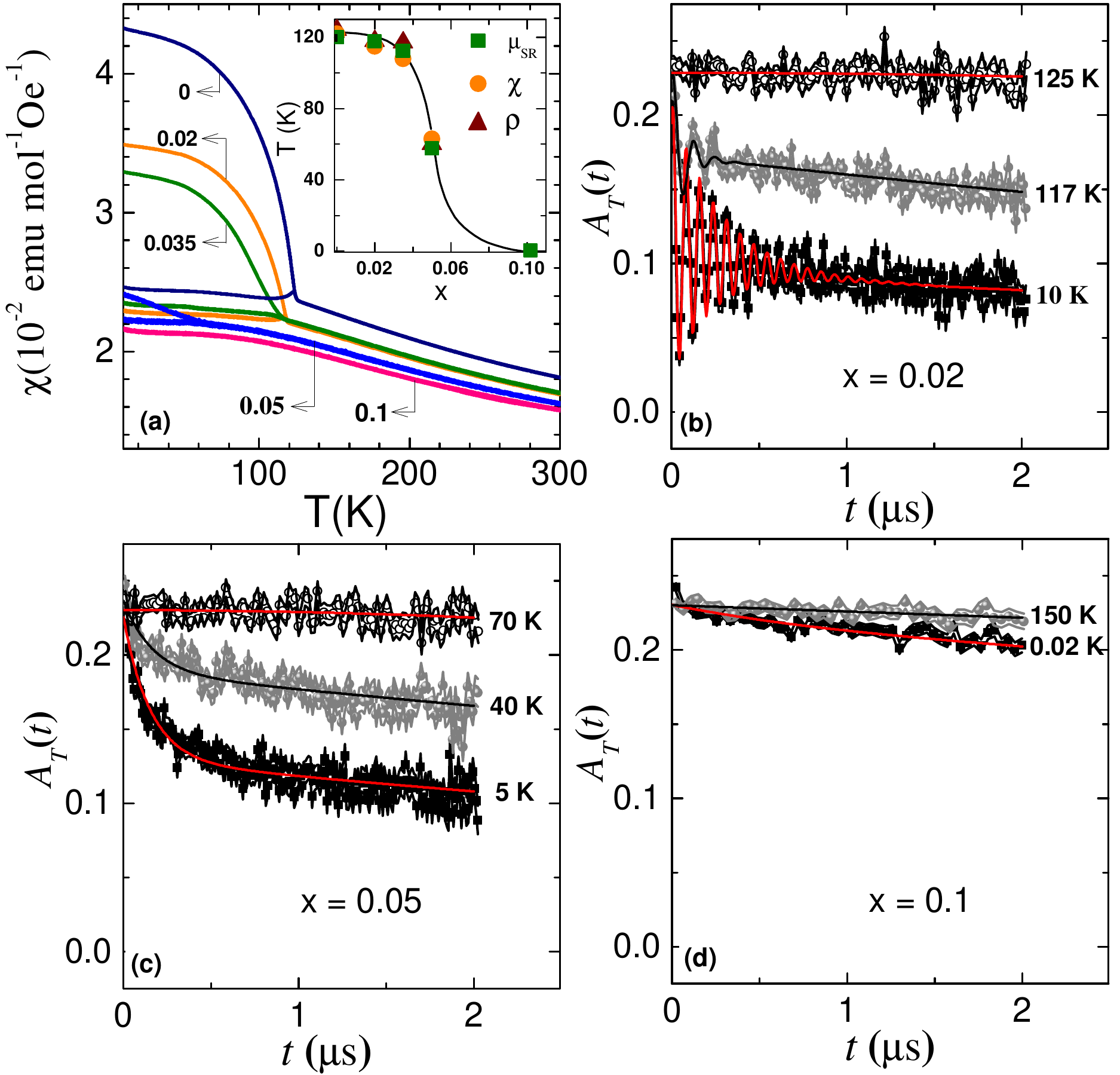}
	\caption{\label{chi} (a) The susceptibility ($\chi$) for $x \leq 0.1$ is shown as a function of temperature. Thermomagnetic hysteresis is seen only for $x = 0, 0.02, 0.035$ and $0.05$ sample; the upper(lower) branch represents FC(ZFC) data. The inset shows variation of $T_{N}$ (from $\chi$ and $\mu^{+}$SR data) and $T_{MI}$ (from $\rho$ data) with doping. The line in the inset is a guide to eye. (b)-(d) Representative zero-field depolarization curves for the $\mu^{+}$ spins in the time domain for $x = 0.02, 0.05$ and $0.1$. The continuous lines are fit to the data (see text for details)}
\end{figure}

In Fig. \ref{chi}a, we show the temperature dependence of $\chi$ for $0 \leq x\leq 0.1$. The transition to the AIAO state is marked by a cusp in the ZFC data below which the ZFC and FC branches bifurcate in agreement with previous reports \cite{Sagayama, Ishikawa, Lafracois2015, Prando1}. We note that in the anomalous doping range, the rate of change of $T$$_N$ is rather slow ($\Delta T_N/T_N \leq 0.05$) and the qualitative behavior of $\chi(T)$ remains unchanged suggesting that the ground state of EIO is preserved at least up to $x = 0.035$. In the crossover region $\Delta T$$_N$ is suppressed sharply, and for $x = 0.1$ no magnetic ordering is observed down to $T = 2$ K.  

In Fig.~\ref{chi} we show the results of $\mu^{+}$SR measurements for compositions $0 < x\leq 0.1$ where we plot representative time-depolarization curves for the $\mu^{+}$ spins in conditions of zero external magnetic field. It is evident that the long-lived coherent oscillations observed at low temperatures for $x = 0.02$ are highly overdamped for $x = 0.05$, which we interpret as the result of a long-range order to short-range order crossover for the magnetic phase \cite{Prando2}. Accordingly, we refer to the fitting procedure discussed in Ref. \cite{Prando2} to analyze our current data. The fitting results are shown in Fig. \ref{chi}. From these fits, we estimate the magnetic volume fraction ($V_{m}$) as $\sim90 \%$ for $ x = 0.02$ and $\sim70 \%$ for $x = 0.05$ at the lowest temperatures. A more dramatic suppression of the AIAO phase is observed for $x = 0.1$ where no sizeable magnetic contribution of electronic origin is observed down to a temperature of $0.02$ K. A marginal increase in the overall relaxation at $T = 0.02$ K may point towards the onset of dynamical processes, possibly from extrinsic phases. We define $T_{N}$ as the highest $T$ value where a transversal relaxation is discernible. The values of $T_N$ based on this criterion agree nicely with those obtained from the ZFC-FC splitting in $\chi(T)$ as shown in Fig.~\ref{chi}a(inset).

\begin{figure}[t!]
	\includegraphics[width = 8.5 cm]{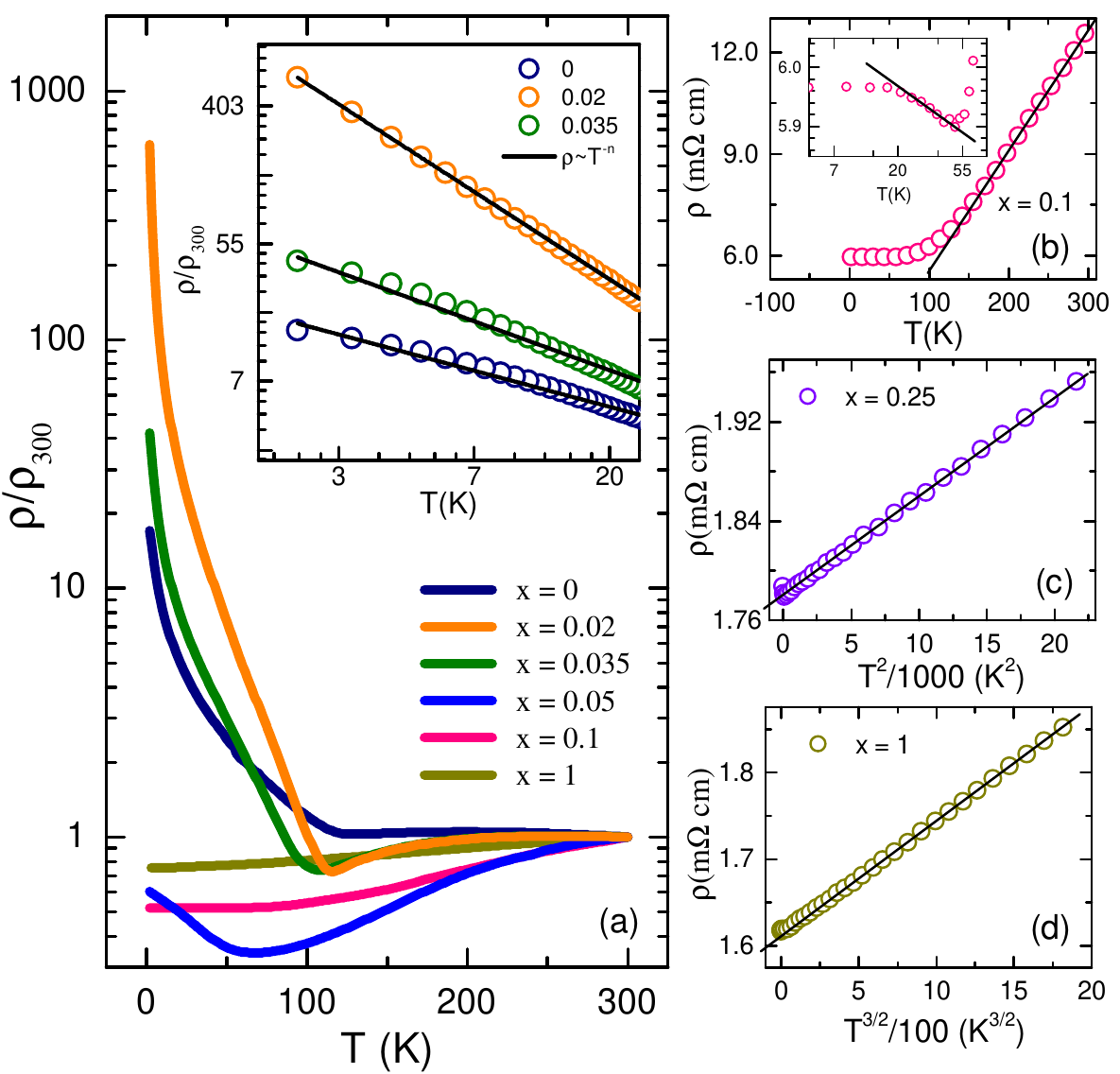}
	\caption{\label{rho} (a) The temperature variation of normalized resistivity $\rho/\rho_{300}$ for various $x$. The inset in (a) shows the power law fit at low-temperatures for $x = 0, 0.02$ and $0.035$. (b), (c) and (d) show the temperature variation of $\rho$ for $x = 0.1, 0.25$ and $1$ samples, respectively. The straight line highlights $T$ (b), $T^2$ (c) and $T^{3/2}$ (d) variation for $x = 0.1, 0.25$ and $1$, respectively. The inset in (b) shows the low-temperature resistivity minimum. The straight line showing a $-lnT$ increase is a guide to eye.}
\end{figure}

The resistivity of our samples is shown in Fig. \ref{rho}. For $x = 0$, $\rho(T)$ exhibits a behavior similar to that previously reported with a MI transition at $T_{MI} = 120$ K, below which $\rho$ increases sharply upon cooling. In concurrence with $\chi$ and $\mu SR$, in the anomalous range, $T_{MI}$ decreases very slowly and the qualitative behavior of $\rho$ remains unchanged. One would expect a slight broadening of the MI transition due to chemical disorder, but surprisingly this transition is sharpest for the $2\%$ Bi doped sample. This is analogous to the effect of external pressure previously reported by Tafti et al. \cite{Tafti} who showed that pressure up to $6$ GPa tends to sharpen the MI transition. This suggests that any broadening of the MI transition due to chemical disorder is masked by the negative chemical pressure. Attempts to fit $\rho$ in this doping range to either the Arrhenius or the variable-range-hopping (VRH) models did not yield satisfactory results. However, a $1/T^\alpha$ power law dependence provides a better description of $\rho(T)$ as shown in the Fig. \ref{rho}a(inset). For $x = 0.02$, this fit is rather satisfactorily up to $T = 20$ K, and with $\alpha \approx 1$; but for $x = 0.035$ and $0$, not only the fitting range shrinks, the value of exponent also reduces to $\alpha \approx 0.5$. In the crossover region, $\rho$ changes dramatically with the MI transition turning broad, centered near $T = 70$ K, consistent with $\chi$ or $\mu_{SR}$ data. Upon entering the normal region, a metallic behavior with a prominent $T^n$ dependence ensues. For $x = 0.25$, the behavior is that of a Fermi-liquid with $T^2$ dependence almost over the whole temperature range (Fig. \ref{rho}c); but this changes to $T^{3/2}$ for $x \ge 0.5$. For BIO, a $T^{3/2}$ dependence is shown (Fig. \ref{rho}d) in agreement with a previous report \cite{Qi}.At the boundary $x = 0.1$ where AIAO $\rightarrow 0$, $\rho(T)$ shows a $T$-linear variation over a broad temperature range (Fig. \ref{rho}b). An expanded view of the low temperature region reveals the presence of an upturn with $\rho$ varying as $-lnT$ down to $T \approx 25$ K below which it tends to saturate (Fig. \ref{rho}b inset).    


\begin{figure}[t!]
	\includegraphics[width =8 cm]{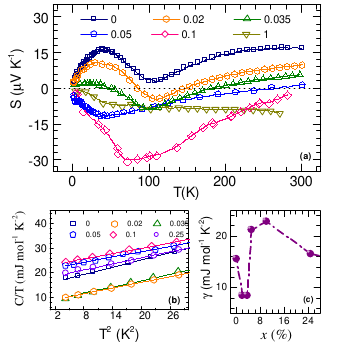}
	\caption{\label{sandgamma} (a) The temperature variation of thermopower $S$ is shown for a few representative samples; (b) the specific heat is plotted as $C_p/T$ versus $T^2$. The straight lines through the data points are linear fits $\gamma + \beta T^2$; (c) the variation of $\gamma$ is shown as a function of $x$.}
\end{figure} 
 
In Fig. \ref{sandgamma}, we show the temperature dependence of the thermopower $S(T)$ for some representative samples. In EIO, $S$ remains positive for all temperatures indicating that the dominant charge carriers are hole-like. The qualitative behavior and sign of $S$ are in good agreement with previous reports \cite{Matsuhira2011, telang}. In the anomalous doping range, while the qualitative form of $S$ remains unchanged, the entire $S(T)$ curve shifts rigidly downward resulting in negative $S$ for intermediate temperatures. In  the crossover range, consistent with $\chi$, $\mu SR$ and $\rho$, $S$ has changed dramatically as shown for $x = 0.05$, with a negative sign of $S$ almost over the whole temperature range. In BIO, $S$ remains negative and shows a typical metal-like behavior. Finally, for $x = 0.1$, $S$ is not only negative but it is also qualitatively different from either of $x = 0$ or $1$, featuring a maximum near $T = 60 K$ ($S_{max} = -30 \mu V K^{-1}$), which is the highest $S$ for any $x$ in this series. From the variation of $S$ with $x$ we infer that the sign of majority charge carrier changes from hole-like ($x \le 0.035$) to electron-like ($x \ge 0.05$). 

In Fig. \ref{sandgamma}, $C_p/T$ is plotted against $T^2$. In the low-temperature range, $C_p/T$ exhibits a good linearity that can be fitted using $C_p = \gamma T + \beta T^3$ where $\gamma$ and $\beta$ represents the electronic and phononic contributions, respectively. The spin wave contribution is expected to be small due to small-size of the ordered Ir moment \cite{Disseler02}. The variation of $\gamma$ with $x$ is shown in Fig. \ref{sandgamma}. For EIO, $\gamma$ of $15$ mJ mol$^{-1}$K$^{-2}$ is in good agreement with previous report \cite{Ishikawa} and suggests moderately strong electronic correlations as predicted theoretically \cite{Krempa2013}. Upon Bi doping, $\gamma$ decreases in the anomalous range before increasing again as $x$ enters the crossover range. For $x = 0.1$, $\gamma$ reaches the highest value of  $\sim25$ mJ mol$^{-1}$K$^{-2}$ before declining again with further increase in $x$.

We now discuss the anomalous lattice contraction. It is surprising that despite its larger size ($r_{Bi^{3+}} =1.17\AA$) compared to Eu ($r_{Eu^{3+}} = 1.066\AA$), initial Bi-doping leads to a lattice contraction. Since BIO has a stable pyrochlore structure with Bi in $+3$ oxidation state, it is unlikely that Bi doped in EIO has a valence state different from $+3$. We nevertheless confirmed this by performing XPS measurements at DESY synchrotron facility \cite{Telang2019}. To put this anomalous behavior into perspective, we should mention here that in the structurally analogous (Y$_{1-x}$Bi$_x$)$_2$Ru$_2$O$_7$ \cite{Kanno} or (Eu$_{1-x}$Sr$_x$)$_2$Ir$_2$O$_7$ series \cite{Banerjee} no such lattice anomaly has been reported, and the MI transition is also shown to suppress gradually.

Though unusual, deviation from the Vegard's law in TMOs is not completely unknown. For example, La doping in SrTiO$_3$ \cite{Janotti}, or in the iridates Sr$_3$Ir$_2$O$_7$ and Sr$_2$IrO$_4$ \cite{Hogan2017,Chen, Hogan2017}, has been shown to result in an anomalous lattice expansion. This is believed be an electronically driven effect arising from carrier doping due to La$^{3+}$ substitution for Sr$^{2+}$. It is argued that upon carrier doping, in some cases, it may be energetically favorable for a system to lowers its total energy by shifting the electronic bands through an expansion or contraction of the lattice, which is known as the deformation potential effect \cite{Bardeen1950}. In our study, we observe that the anomalous lattice contraction preserves the magnetic and insulating ground state which is suppressed as the lattice expands, which makes it evident that this anomaly is indeed intimately tied up to the electronic properties. However, here a notable exception arises from the isovalent nature of Eu$^{3+}$ and Bi$^{3+}$, which precludes the conventional carrier doping mechanism. How then are carriers doped when Bi is substituted for Eu? In a recent study, Qi et al.  \cite{Qi} found that in the Bi doped pyrochlores, the Ir($5d$)--(Bi)$6s$/$6p$ hybridization can be significantly enhanced, which makes the $6s/6p$ electrons of Bi to contribute to the DOS at $\epsilon_F$.    

This brings us to the point concerning to the robust ground state of EIO against Bi doping. As shown here, during initial Bi doping the AIAO/MI transitions become sharper (sharpest for $x = 0.02$), and low-temperature $\rho$ approaches a $1/T$ behavior. We suggest that these are manifestations of the WSM phase that has been concretely established in the theoretical works. We conjecture that $\epsilon_F$ at $x = 0$ is located slightly below the Weyl nodes (positive $S$ for $x = 0$); and, upon initial Bi-doping, lattice contracts to shift the bands down (deformation potential effect), which pushes $\epsilon_F$ up, closer to the Weyl nodes. This is consistent with the decreasing behavior of $\gamma$ which is proportional to the DOS at $\epsilon_F$. This also explains why $\rho(T)$ at low-temperatures approaches a $1/T$ behavior. In the WSM phase, the electron-hole symmetry about the Weyl node results in current carrying states with \textit{zero} total momentum; the electron-electron interaction therefore relaxes the current but with zero momentum transfer leading to a $1/T$ dependence \cite{Hosur}. Indeed, as $x \rightarrow 0.02$, the low-temperature $\rho(T)$ is closest to the expected behavior. At higher Bi doping, $\epsilon_F$ is gradually tuned away from the Weyl nodes, which explains the changing sign of $S$. This picture, however, breaks down as $x$ increases further due to suppression of the AIAO state. 

For $x = 0.1$, the lack of magnetic ordering, a large $\gamma$ value, and a resistivity minimum showing a $-lnT$ dependence at low-T are all reminiscent of Pr$_2$Ir$_2$O$_7$ (PIO), which features a quadratic band touching point protected by the inversion and time reversal symmetries \cite{NakatsujiPr227, Kondo2015}. Recently, the resistivity minimum in PIO has gathered a renewed interest \cite{Cheng, Kondo2015, Udagawa}. In the earlier studies, it was interpreted as arising due to the Kondo effect from localized Pr moments, but it is now believed to be intrinsic to the Ir subsystem, which has the characteristics of a 3D Luttinger semimetal with interactions \cite{Moon, Cheng}. From this point of view, $x = 0.1$ or (Eu$_{0.9}$Bi$_{0.1}$)$_2$Ir$_2$O$_7$ sample is interesting as it mitigates the complexity arising due to a magnetic $A$-site. One can then argue that (Eu$_{0.9}$Bi$_{0.1}$)$_2$Ir$_2$O$_7$ is possibly the mother phase featuring a quadratic band touching point analogous to PIO from which the WSM phase can be derived in the TRS broken regime close to $x = 0$; and other exotic, but not necessarily topologically non-trivial, electronic phases appear for $x > 0.1$. In particular, the region from $0.1 < x < 1$ appears to be very fertile as the $\rho(T)$ behavior changes from $T$-linear to $T^{3/2}$, and with a Fermi-liquid like $T^2$ behavior near $x = 0.25$. This region should be explored in more details in future.   

To summarize, we show that Bi-doping in EIO provides a unique platform, which, unlike Sr or Ca doping, does not create a Ir$^{4+}$ -- Ir$^{5+}$ charge disproportionation, and thus keep the Ir$^{4+}$--sublattice intact which is key to obtaining various non-trivial topological phases. We report here three important findings: (i) anomalous lattice contraction strongly tied to the electronic properties; (ii) robust ground state of EIO against initial Bi doping, which we attribute the to the topological WSM phase, which can be realized for very small Bi-doping which pushes $\epsilon_F$ closer the Weyl node without destroying the linear dispersion; (iii) (Eu$_{0.9}$Bi$_{0.1}$)$_2$Ir$_2$O$_7$, located at the boundary from which other non-trivial topological phases can be derived. Our findings are expected to motivate further research in exploring new quantum phases in the $U - \lambda$ phase space of $5d$ transition metal oxides. 

\
\begin{acknowledgments}
SS would like to thank the Department of Science and Technology(DST) and Science and Engineering Research Board (SERB), India for financial support under grants No. EMR/2014/000365. P. T. and S. S. acknowledge DST for travel grant for performing synchrotron x-ray and $\mu$SR experiments. G. P. acknowledges the kind support of the Laboratory for Muon Spin Spectroscopy at the Paul Scherrer Institute during the muSR experiments. AKS acknowledges DST for financial support. 
\end{acknowledgments}

\bibliographystyle{apsrev4-1}
\bibliography{EuBi}

\end{document}